\documentclass[aps,twocolumn,amsmath,amssymb,showpacs,superscriptaddress,floatfix]{revtex4-1}
\usepackage{graphicx}
\begin{document}
\title{Entropic force of polymers on a cone tip}

\author{Mohammad F. Maghrebi}
\email{Email: magrebi@mit.edu}
\affiliation{Massachusetts Institute of Technology, Department of
  Physics, Cambridge, Massachusetts 02139, USA}
\author{Yacov Kantor}
\affiliation{Raymond and Beverly Sackler School of Physics and
Astronomy, Tel Aviv University, Tel Aviv 69978, Israel}

\author{Mehran Kardar}
\affiliation{Massachusetts Institute of Technology, Department of
  Physics, Cambridge, Massachusetts 02139, USA}

\date{\today}


\begin{abstract}
We consider polymers attached to the tip of a cone, and the resulting force due to
entropy loss on approaching a plate (or another cone). At separations shorter than
the polymer radius of gyration $R_{g}$, the only relevant length scale is the tip-plate
(or tip-tip) separation $h$, and the entropic force is given by $F={\cal A} \, k_{B}T/h$.
The universal amplitude $\cal A$ can be related to (geometry dependent) correlation
exponents of long polymers. We compute $\cal A$ for phantom polymers, and for
self-avoiding (including star) polymers by $\epsilon$-expansion, as well as by
numerical simulations in 3 dimensions.
\end{abstract}

\maketitle
Single molecule manipulation \cite{bustamante2003,kellermayer,neuman,deniz,neuman2008}
using techniques such as atomic force microscopy (AFM) \cite{fisher},
microneedles \cite{kishino}, optical \cite{neuman2004,hormeno} and
magnetic \cite{gosse} tweezers enable extremely detailed study of geometry
and forces in long polymers. The positional accuracy of AFM tip
\cite{kikuchi1997,neuman2008} can be as good as few nm, while the forces of order of
1 pN can be measured, and measurements can be carried out in almost biological
conditions \cite{brk1997,drake1989}. These enhanced sensitivities bring us to the
range where entropic forces of long polymers in a solvent can be significant even
when the deformation of the polymer is relatively slight. While the main thrust
of the experimental research is extraction of specific information from the
force-displacement behaviors of the polymers, certain features are independent
of the microscopic details \cite{degennesSC}, but depend on the probe shape,
as discussed in this work.

We consider an idealized set-up in which a polymer is attached to the tip of a
solid cone. The cone approaching a plate (or another cone) exemplifies a geometry
in which the only (non-microscopic) length scale is provided by the tip-plate
(or tip-tip) separation $h$. Fluctuating polymers  typify self-similar variations
at scales intermediate between microscopic (persistence length $a$) and macroscopic.
The latter is set by the radius of gyration which grows with the number of monomers
through the scaling relation $R_{g}\propto N^\nu$. Thus when a cone-tip-attached
polymer approaches a plate, at separations $a\ll h\ll R_{g}$ the only relevant
length scale is $h$, and on dimensional grounds, the force due to loss of entropy
must behave as
\begin{equation}
\label{Eq:force}
F={\cal A}\,\frac{k_BT}{h}\ .
\end{equation}
Such a force law should apply to all circumstances where the separation provides
the only relevant length scale.  The amplitude $\cal A$ will depend on geometric
factors such as the opening angle of the cone $\Theta$ (and if tilted, on the
corresponding angle). One could presume, that the dimensionless amplitude may
also depend on microscopic properties on the polymer. However, in case
of {cone-tip-}polymers we shall demonstrate that the amplitude $\cal A$ can be related to
universal (and shape dependent) polymer exponents.
The simple force law of Eq.~(\ref{Eq:force}) follows easily from various polymer
scaling forms (see, e.g. the derivation below) such as in
Refs.~\cite{Eisenriegler1982,Duplantier1986,Rowghanian2011}, and should be part of polymer lore.
Surprisingly, we could not find an explicit reference to it in any of the standard polymer
textbooks.

\begin{figure}
\includegraphics[width=8cm]{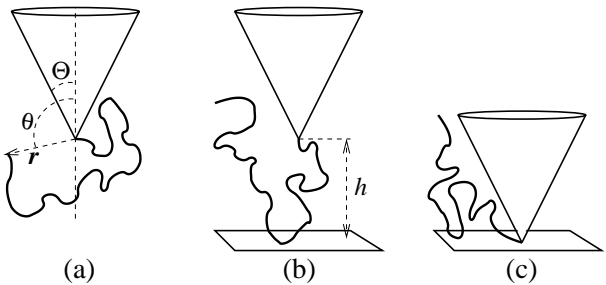}
\caption{(a) Polymer attached to the tip of a solid  cone with apex
semi-angle $\Theta$ (configuration ``c"); positions are described by the spherical
coordinates $r,\theta$ and azimuthal angle $\phi$ (not shown).
(b) The tip is at a distance $h\ll R_g$ from the plate.
(c) The tip touching the plate (configuration ``cp").
\label{fig:def_geometry}}
\end{figure}

A polymer attached to the tip of an AFM is approximated
as linked to the apex of a cone as depicted in Fig.~\ref{fig:def_geometry}a.
With the cone far away from a plate ($h\gg R_{g}$), the number of configurations of
the polymer grows with the number of monomers as
\begin{equation}\label{Eq:gammadef}
{\cal N}_{\rm c}= b \, z^N  N^{\gamma_{\rm c}(\Theta)-1},
\end{equation}
where the effective coordination number $z$, as well as the pre-factor $b$,
depend on the microscopic details, while the  `universal' exponent $\gamma_{\rm c}$  only
depends on the cone angle.
When the cone touches the plate as in Fig.~\ref{fig:def_geometry}c, the number of
configurations is reduced to ${\cal N}_{\rm cp}$ with the same form as Eq.~\eqref{Eq:gammadef},
but with a different exponent $\gamma_{\rm cp}(\Theta)$.
We shall henceforth use the exponent subscript `${\rm s}$' (as in $\gamma_{\rm s}$)
to refer to the above cases, with ``s=c" for cone and ``s=cp" for cone+plate;
the absence of a subscript (as in $\gamma$) will signify a free polymer.
The work done against the entropic force in bringing in the tip from
afar to contact the plate can now be computed  from Eq.~\eqref{Eq:force} as
\begin{equation}
\label{Eq:work}
W=\!\int_a^{R_g}\!\!{\rm d}h~{\cal A}\frac{k_BT}{h}\ =\!{\cal A}k_BT\ln\frac{R_g}{a}
 ={\cal A}\,\nu k_BT\ln N.
\end{equation}
The work can also be computed from the change in free energies between the final
and initial states, due to the change in entropy, as
\begin{equation}
\label{Eq:DF}
\Delta {\cal F}= -T\Delta{\cal S}=T{\cal S}_{\rm c}-T{\cal S}_{\rm cp}
=k_BT(\gamma_{\rm c}-\gamma_{\rm cp})\ln N  \,,
\end{equation}
where the entropy ${\cal S}=-k_B\ln{\cal N}$ was computed from Eq.~(\ref{Eq:gammadef}).
By equating $W$ and $\Delta {\cal F}$ we find
\begin{equation}
\label{Eq:A}
{\cal A}=\frac{\gamma_{\rm c}-\gamma_{\rm cp}}{\nu}=\eta_{\rm cp}-\eta_{\rm c}\, ;
\end{equation}
the final result obtained from the scaling law $\gamma_{\rm s}=(2-\eta_{\rm s})\nu$,
where $\eta$ characterizes the anomalous decay of correlations ($\sim 1/r^{d-2+\eta}$).

 {In the above discussion we have assumed that the only interaction between
 the polymer and the surfaces is due to hard-core exclusion.
 Attractive interactions between the polymer and surface will introduce
 temperature dependent corrections, and an additional size scale.
Weak interactions are asymptotically irrelevant, but strong interactions may
lead to a phase in which the polymer is absorbed to the surface, where
the entropic considerations presented here will no longer be appropriate.}

We {have derived} Eq.~\eqref{Eq:DF} for the specific case of a cone and a plate. However, this equation can
be applied to any situation where in the limits of large and vanishing separations
we have scale-free shapes. In particular, the same rule can be applied when a linear or
star polymer is brought from infinity to a contact with a repulsive plane, since both
extremes have no length-scale, leading to exact expressions for the force
constant \cite{Duplantier1986}. However, if the AFM tip is slightly rounded,
an additional length scale is introduced, and one may expect {a} non-trivial
crossover between various regimes \cite{bubis}.

We have thus reduced the computation of the force  to calculation
of correlation functions. In the absence of self-avoidance and other interactions,
correlations of the so-called ideal or phantom polymer (henceforth denoted by
subscript 0) are the same as a free-field
theory, satisfying (at scales  shorter than $R_g$)
the Laplace equation $\nabla^2 G{_0}({\bf r}, {\bf r'})=-\delta({\bf r}-{\bf r'})$ \cite{degennesSC}.
With one point at a short distance $a$ from the cone, correlations behave as
$G{_0}(a,{\bf r})\sim a^{\eta{_0}}/r^{d-2+\eta{_0}}\Psi(\theta)$, where $r$ and $\theta$
denote the distance of ${\bf r}$ from the tip, and the angle of ${\bf r}$ to the
cone axis.
{The change in scaling
from a free phantom polymer is captured by $(a/r)^{\eta_0}$,}
{and $\Psi(\theta)$ is a dimensionless function depending only on the polar angle due to the symmetry of the geometry.} (We  consider  a generalized cone in $d$ spatial dimensions characterized
by a single polar angle $\theta$, and $d-2$ azimuthal angles $\phi, \psi, \cdots$.)
Substituting the above form in the Laplace
equation, we find that the exponent $\eta{_0}$ satisfies
\begin{equation}
\label{Eq:psi}
\frac{1}{(\sin\theta)^{d-2}}\frac{d}{d \theta}\left[(\sin\theta)^{d-2}
\frac{d\Psi}{d  \theta}\right]+\eta{_0}(d-2+\eta{_0})\Psi(\theta)=0\,,
\end{equation}
with an appropriate boundary condition on $\Psi$. For an isolated cone,
the function $\Psi$ must be positive and regular outside the cone,
with  $d\Psi/d\theta|_{\theta=\pi}=0$ to avoid a cusp on the
symmetry axis, and $\Psi(\Theta)=0$ on the cone surface.
For the cone+plate, the appropriate solution is positive  and vanishes
both at $\theta=\Theta$ and $\theta=\pi/2$. The first case was considered
by Ben-Naim and Krapivsky~\cite{BenNaim} in connection with diffusion near
an absorbing boundary~\cite{redner_book}, and we follow these derivations.
The solution in general $d$ requires the use of associate{d} Legendre
functions, but simplifies in a few cases described below.

\begin{figure}
\null\vskip 1cm
\includegraphics[width=7.5cm]{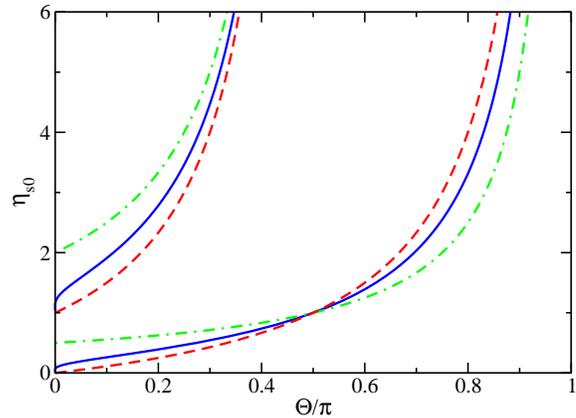}
\caption{
\label{fig:ThAll}
(Color online) The exponent $\eta_{\rm s0}$ for ideal polymers in $d=$2 (dot-dashed),
3 (solid), 4 (dashed) for cone (``s=c") of angle $\Theta$ (bottom curves), and
``s=cp" (top curves).
}
\end{figure}

\par\noindent$\bullet$
{\it For $d=2$,} Eq.~\eqref{Eq:psi} reduces to
$\Psi''+\eta{_0}^2\Psi=0$; solved by  linear combinations of
$\sin(\eta{_0}\theta)$ and $\cos(\eta{_0}\theta)$ to yield
\begin{equation}
\label{Eq:beta2d}
\eta_{\rm c{0}}=\frac{\pi}{2(\pi-\Theta)},\quad{\rm and}\quad
\eta_{\rm cp{0}}=\frac{2\pi}{\pi-2\Theta}\,.
\end{equation}
Both results (depicted in Fig.~\ref{fig:ThAll}) go to a finite value as
$\Theta\to0$, reflecting the strong reduction in configurations due
to the remnant (barrier) line, and $\eta{_0}\to\infty$ when
the boundaries confine the polymer to a vanishing sector.

\par\noindent$\bullet$
{\it For $d=4$,} the substitution $\Psi=u/\sin\theta$ simplifies Eq.~\eqref{Eq:psi} to
$u''+(\eta{_0}+1)^2u=0$, solved by a linear combination of $\sin[(\eta{_0}+1)\theta]$ and
$\cos[(\eta{_0}+1)\theta]$,
and we find
\begin{equation}
\label{Eq:beta4d}
\eta_{\rm c 0}=\frac{\Theta}{\pi-\Theta},\quad{\rm and}\quad
\eta_{\rm cp 0}=\frac{\pi+2\Theta}{\pi-2\Theta}\, ;
\end{equation}
depicted by the bottom and top dashed lines in Fig.~\ref{fig:ThAll}. {The cone
exponent $\eta_{ {\rm c}0}$ vanishes linearly with $\Theta$---a needle in four dimensions is invisible.}

\par\noindent$\bullet$
{\it For $d=3$,} Eq.~\eqref{Eq:psi} is solved by a linear combination
of regular Legendre functions
$P_{\eta{_0}}(\cos\theta)$ and $Q_{\eta{_0}}(\cos\theta)$.
The resulting exponents (which cannot be cast as simple functions),
are plotted as solid lines in Fig.~\ref{fig:ThAll}. {The exponent $\eta_{{\rm c}0}(\Theta)$ vanishes with the angle, but b}oth curves
approach their limiting value as $\Theta\to0$ with infinite slope
{via} a logarithmic singularity ($\sim1/|\ln\Theta|)$.

\par\noindent$\bullet$
{\it For all $d$,} the cone becomes a plate for $\Theta=\pi/2$.
Correlations with one point approaching a surface are easily obtained by
the method of images \cite{binder} leading to $\eta_{\rm c{0}}=1$, which is
clearly seen in Fig.~\ref{fig:ThAll}. For $3<d<4$,  both exponents approach
their limiting value when $\Theta\to 0$ as $\Theta^{p_0}$ with $p_0=d-3${:}
{
\begin{align}\label{Eq. eta0 needle}
  \eta_{\rm c0} &= \frac{\Gamma(1-\epsilon/2)}{\sqrt{\pi}\, \Gamma(1/2-\epsilon/2)}\Theta^{1-\epsilon}, \qquad{\rm and} \nonumber \\
  \eta_{\rm cp0} &= 1+\frac{4\Gamma(2-\epsilon/2)}{\sqrt{\pi}\, \Gamma(1/2-\epsilon/2)}\Theta^{1-\epsilon},
\end{align}
where $d=4-\epsilon$. These equations}
reflect the fact that the two dimensional phantom polymer will not intersect
the remnant line in $d>3$. For $d<3$, the limiting value is different from
the case without any cone indicating the finite probability of intersection of
the polymer with the line.

Universal aspects of swollen (coil) polymers with short-range interactions
can be modeled by a self-avoiding walk (SAW). In $d=3$,  a SAW in empty space
has  exponent $\gamma\approx1.158$ \cite{caracciolo}, while in $d=2$,
$\gamma=43/32$~\cite{madras_book} (for ideal polymers $\gamma=1$ at any $d$).
There are a number of results regarding $\gamma_{\rm s}$
for polymers confined by wedges in 2D and
3D~\cite{cardy_red,cardy1,cardy2,guttmann,gsurface,debell}: A SAW
anchored at the origin and confined to a solid wedge (in 3D) or a
planar wedge (in 2D) has an angle-dependent $\gamma_{\rm s}$ that diverges as
the confining angle vanishes.  Studies of a polymer attached to the tip of a 2D
sector in 3D, and to the apex of a cone have been performed \cite{slutsky}.
Extensive analytical \cite{kosmas,douglas_kosmas} and numerical
\cite{gsurface,debell} studies of SAWs anchored to a
solid plate in 3D find $\gamma_{\rm s} \equiv \gamma_1\approx
0.70$ \cite{gsurface}, or 0.68 \cite{grass_gamma1}. For ideal polymers
$\gamma_1=1/2$ for any $d$. We are unaware of specific results for the
geometries depicted in Fig.~\ref{fig:def_geometry},
and report below our numerical and analytical estimates.

\begin{figure}
\null\vskip 1cm
\includegraphics[width=8cm]{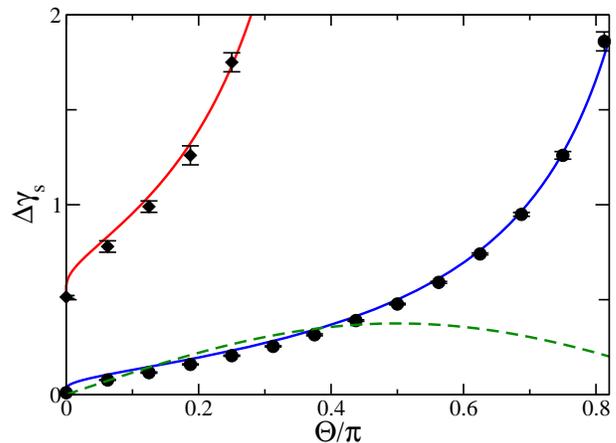}
\caption{
\label{fig:dgammavsalpha}
(Color online) Dependence of the exponent difference
$\Delta\gamma_s=\gamma-\gamma_{\rm s}$ on angle $\Theta$
for the  cone+plate (top curves), and an isolated cone (bottom curves).
The solid curves are the results for an ideal polymer, while
the data points represent numerical results for SAWs in the same geometry.
All error bars are estimates of the $N\to\infty$ extrapolation error.
The dashed line depicts prediction of the $\epsilon$-expansion in Ref.~\cite{slutsky}.
}
\end{figure}

We performed numerical simulations for SAWs of lengths $N=16$, 32, \dots, 1024
on a cubic lattice. The SAWs were generated by a dimerization method
\cite{dimerization1,dimerization2}
in which an unbiased $N$-step SAW is created by attempting to join
two $N/2$-step SAWs previously obtained by the same algorithm.
We generated $10^8$ SAWs for each $N$, each of which is attached
to the origin and checked whether it touches a cone (or cone+plate).
The probability that an $N$-step SAW does not intersect the excluded space
is the ratio of permitted number of walks to the total number of SAWs, i.e.
$p_N={\cal N}_{\rm s}/{\cal N}\sim N^{\gamma_{\rm s}-1}/N^{\gamma-1}=N^{\Delta\gamma_{\rm s}}$.
The ratio $p_N/p_{2N}=2^{\Delta\gamma_{\rm s}}$ is then used to estimate
$\Delta\gamma_{\rm s}$ for each sequential pair of $N$s.
The results are extrapolated to $N\to\infty$ by  plotting the estimates {\em versus}
$1/\sqrt{N}$; errors are due to both  finite sample size and insufficiently large $N$.
The estimated exponents (full symbols in Fig.~\ref{fig:dgammavsalpha}) are rather
close to the results for ideal polymers. This apparent proximity is somewhat
misleading  as $\Delta\gamma_{\rm s}=\gamma-\gamma_{\rm s}$
masks an appreciable shift in both $\gamma$ and $\gamma_{\rm s}$.
For example, for a SAW near an isolated cone we expect
$\Delta\gamma(\Theta=\pi/2)=\gamma-\gamma_1=$0.44-0.48,
very close to ideal polymer result of 1/2; our numerical estimate at
$\Theta=\pi/2$ is $0.477\pm0.004$.

The dashed line in Fig.~\ref{fig:dgammavsalpha} depicts the result~\cite{slutsky}
of an $\epsilon=(4-d)$ expansion in which the 2-dimensional {\em surface of the cone}
is treated as a weakly repulsive potential.
The lowest order result, $\Delta\gamma_{\rm c}=(3\epsilon/8)\sin\Theta$,
captures some features of the numerics for $\Theta<\pi/2$, but fails dismally
for  $\Theta>\pi/2$ (since the polymer simply jumps to the larger space inside
 the cone). It also fails to capture the correct behavior
as $\Theta\to0$. A better approach is to exclude the  {\em interior of
the cone}, and to this end we follow the work of Cardy~\cite{cardy1,cardy2}
for the wedge geometry. The corresponding computations for a cone are  more
complicated, and we relied upon recent results on
the electrodynamic Casimir force on a cone~\cite{Maghrebi10}.

For an $\epsilon$-expansion we need the full non-interacting Green's function in
4-dimensions, going beyond the large separation asymptotic form obtained via
Eq.~(\ref{Eq:psi}). ({{{For a general $\epsilon$-expansion, one has to know the Green's function in $4-\epsilon$ dimensions, however, we carry out this expansion only to first order in $\epsilon$. Since the strength of the interaction is linear in $\epsilon$, it suffices to obtain the Green's function in 4-dimensions.}}}) Following Ref.~\cite{Maghrebi10}, this is given by
\begin{eqnarray}
&\!\!G_0(x,x')= \!\! {\sum\atop {klm} }
\frac{ (-1)^l\pi}{2}  \frac{\Gamma(\rho_k+l+1)}{\sin(\rho_k \pi) \Gamma(\rho_k-l)}
\frac{P_{\rho_k-1/2}^{-l-1/2}(\cos\Theta)}{\partial_{\rho_k}P_{\rho_k-1/2}^{-l-1/2}(-\cos\Theta)}
\nonumber \\
&\!\!\frac{r_<^{\rho_k-1}}{r_>^{\rho_k+1}}
\frac{P_{\rho_k-1/2}^{-l-1/2}(-\cos\theta)Y_{lm}(\psi, \phi)}{\sqrt{\sin\theta}}
\frac{P_{\rho_k-1/2}^{-l-1/2}(-\cos\theta')Y^{\star}_{lm}(\psi', \phi')}{\sqrt{\sin\theta'}}
,\quad
\label{Eq:G0}
\end{eqnarray}
where the sum is over the triplet of integers $k>0,~l\geq 0$, and $-l\leq m \leq +l$.
The important exponent $\rho_k$ is  the $k$-th root of the transcendental equation
$
  P_{\rho_k-1/2}^{-l-1/2}(-\cos\Theta)=0.
$
{This Green's function is broken up in radii ($r^<$ and $r^>$), as appropriate
to a polymer in the presence of a cone with one endpoint close to the tip and the other end far away.}

The interacting Green's function at first order is obtained by subtracting polymer configurations
that self-intersect, forming an intermediate loop, as
\begin{equation}
G_1=
G_0
\! -u\!\!\int \!{\rm d}^4{ x''}\! G_0(x,x'')
G_0^r(x''\!,x'')  G_0(x''\!,x').
\label{Eq:G1}
\end{equation}
In the language of quantum field theory, the first term is the ``free'' Green's function, albeit in the presence of external boundary conditions, while the second term is the one-loop correction. Note that the intermediate Green's function is regularized by subtracting the Green's function in empty space, hence the superscript $r$.
In the above equation, $u$ is the strength of the self-avoiding interaction, which
is ultimately  set to its fixed point value of $u^*=2\pi^2\epsilon$~\cite{slutsky,cardy1,cardy2}.
To calculate the scaling properties for $r\gg a$, it is sufficient to include only
the first term ($k=1, l=m=0$) from Eq.~\eqref{Eq:G0} in the non--loop propagators
(as in the wedge computation by Cardy~\cite{cardy1,cardy2}, higher-order terms give corrections in higher powers of $a/r$).
The intermediate loop, however, can be of any size requiring the entire sum
(albeit regularized by subtracting the result for $\Theta=0$ to remove an
unrelated divergence). {The loop correction is an integral over the whole space.  The integration over angles $\phi''$ and $\psi''$ is trivial due to  symmetry and yields $l(l+1)$ when summed over all spherical harmonic functions of degree $l$; in the integration over the radius $r''$
we seek a logarithmic contribution which is exponentiated in the end.
Similar to Cardy's analysis, such a logarithm comes only from the region $r<r''<r'$.
There remains an integral over the polar angle $\theta''$,}
but this is cumbersome {as we have to find the roots to the transcendental equation noted before.}
{W}e shall not dwell further on  the {$\epsilon$-expansion for a general opening angle}.
Instead, we focus on the limit of a sharp cone as $\Theta\to 0$, where the leading
singularity  comes {only} from the $l=0$ channel of the loop Green's function. {This is because a sharp cone couples only to the lowest spherical partial wave. }
{A} {c}areful {integral over $\theta''$ and summation over all $k$'s} gives the radial dependence of the Green's function as
\begin{equation}
G_1\propto \frac{r^{\eta_{\rm c 0}}}{{r'}^{\eta_{\rm c0} +2}}\left[1+ \epsilon\ln\frac{r}{r'}\left(\frac{1}{4\pi}\Theta\ln\Theta+.16\, \Theta\right)\right],
\end{equation}
where the term proportional to $\epsilon$ in the bracket is the one-loop
correction to the Green's function. In $d=4-\epsilon$ dimensions,
$\eta_{\rm c0}$ is {given by} Eq.~\eqref{Eq. eta0 needle}{.}
{Now we can exponentiate the radial logarithm to obtain the
renormalized exponent,}
\begin{equation}\label{Eq:Deta1}
  \eta_{\rm c}=\eta_{\rm c0} + \left(\frac{\Theta\ln\Theta}{4\pi}+.16 \, \Theta\right) \, \epsilon.
\end{equation}
Note that the loop-correction vanishes logarithmically with the angle,
as there is no first order in $\epsilon$ contribution to $\eta$ in empty space.
Repeating the same  procedures  for the cone-plate
we find, as $\Theta\to0$,
\begin{eqnarray}\label{Eq:Deta2}
\eta_{\rm cp}=\eta_{{\rm cp}0}+\left(-\frac{1}{8}+\frac{\Theta\ln\Theta}{\pi} +.66\,\Theta\right)\, \epsilon \,{,}
\end{eqnarray}
where the exponent $\eta_{\rm cp0}$ is given by Eq.~\eqref{Eq. eta0 needle}. In this case, the loop-correction goes to $-\epsilon/8$ for $\Theta=0$ due to the presence of the plate.
Equations \eqref{Eq:Deta1} and \eqref{Eq:Deta2}  suggest interpreting the logarithmic
corrections as signatures of a power-law  for $\Theta\to0$. Expanding {$\eta_{\rm c0}$ and $\eta_{\rm cp 0}$}
to the first order in $\epsilon$, we find another  $\ln\Theta$, originating
from the expansion in $4-\epsilon$ dimensions of the phantom polymer (as
opposed to the perturbative terms from the one-loop computation). Summing both
contributions, we obtain
\begin{align}
  \eta_{\rm c}&= \frac{\Theta}{\pi}\left(1-\frac{3}{4}\epsilon\ln\Theta-.06\epsilon\right), \nonumber \qquad {\rm and} \\
  \eta_{\rm cp}&= 1-\frac{\epsilon}{8}+\frac{4\Theta}{\pi}\left(1-\frac{3}{4}\epsilon\ln\Theta-.86\epsilon\right)\,.
\end{align}
We can then recast the approach of the exponents to their
limiting values as a power-law $\Theta^p$ with $p=1-3\epsilon/4$ in place of
$p_0=d-3=1-\epsilon$ for the phantom polymer (Fig.~\ref{fig:ThAll}). We may
interpret this result as follows: self-avoidance swells the polymer, reducing its
fractal dimension from 2 to $\nu^{-1}=2- \epsilon/4$ to lowest order. The
dimensionality of the intersection of such an object with the remnant line is
$d-1-\nu^{-1}=1-3\epsilon/4=p$. We leave this observation as a conjecture for
future studies.

From a practical point of view, the amplitude ${\cal A}=\eta_{\rm cp}-\eta_{\rm c}$
is typically a number of order unity. Thus the  force in Eq.~\eqref{Eq:force}
at room temperature is roughly around 0.1 pN at 0.1 $\mu$m separation;
at the margins of possible measurements by current force apparatus.
The force can be increased by attaching several polymers to the tip. For
ideal (phantom) polymers the force is enhanced by $f$, the number of polymers,
while interactions  modify this conclusion. {The interactions are of two kinds: self-interaction of a single arm (intra-arm) and interactions between two different arms (inter-arm). The  former contribution is computed above to first order in $\epsilon$ and should be multiplied by $f$ for an $f$-arm polymer.
The latter, however, gives rise to a new term multiplied by $f(f-1)/2$,
the number of pairs.
Thus to first order in $\epsilon$, $\eta=f\eta_0+f\eta_i+\frac{f(f-1)}{2}\eta_e$ where $\eta_0$ denotes the exponent of a single phantom polymer, and $\eta_i$ and $\eta_e$ are those of intra-arm and inter-arm interactions respectively.}   In the absence of a cone, the set-up
is similar to the widely studied case of star polymers~\cite{Ohno1989}.
We  carried  the corresponding $\epsilon$-expansion with cone and cone+plate{.} {Here, we just quote the final result after appropriate exponentiations. We} find (in the limit  of sharp cones), a force amplitude
\begin{equation}
\label{starA}
\frac{{\cal A}(f)}{f}=1-\frac{\epsilon}{8}+
\left[\frac{3}{\pi}-\left(.80+\frac{11}{12\pi}(f-1)\right)\epsilon\right]\Theta^{1-3\epsilon/4}\,.
\end{equation}
Interactions amongst the polymers thus {\it reduce} the force amplitude per polymer.

Note that the same exponent $p=1-3\epsilon/4$ governs the approach to a finite
limit in Eq.(\ref{starA}) as $\Theta\to0$.
Another experimental set-up is a cone-tip-attached polymer approaching another cone. The entropic force is of course much
smaller in this case, and we have verified that it also vanishes with cone angle
as $\Theta^p$, bolstering the conjecture that this is a universal exponent related
to the needle geometry.

In summary, we propose that polymers exert an entropic force ${\cal A}k_BT/h$
on a cone tip, with a `universal' amplitude $\cal A$ dependent on geometry,
interactions, and number of polymers. We  conjecture that the singular
form of the amplitude on vanishing cone angle is described by a new exponent.
There are many  set-ups  where a similar force law is expected on the basis
of scaling at length scales shorter than an appropriate correlation length.

\begin{acknowledgments}
{This work was supported by the National Science Foundation under Grants No. DMR-08-03315 (MK, MFM), and  PHY05-51164 (MK at KITP). Y.K. acknowledges the support of Israel Science Foundation grant 99/08.}
The authors acknowledge discussions with B. Duplantier and A. Grosberg.
\end{acknowledgments}


\begin{thebibliography}{37}%
\makeatletter
\providecommand \@ifxundefined [1]{%
 \@ifx{#1\undefined}
}%
\providecommand \@ifnum [1]{%
 \ifnum #1\expandafter \@firstoftwo
 \else \expandafter \@secondoftwo
 \fi
}%
\providecommand \@ifx [1]{%
 \ifx #1\expandafter \@firstoftwo
 \else \expandafter \@secondoftwo
 \fi
}%
\providecommand \natexlab [1]{#1}%
\providecommand \enquote  [1]{``#1''}%
\providecommand \bibnamefont  [1]{#1}%
\providecommand \bibfnamefont [1]{#1}%
\providecommand \citenamefont [1]{#1}%
\providecommand \href@noop [0]{\@secondoftwo}%
\providecommand \href [0]{\begingroup \@sanitize@url \@href}%
\providecommand \@href[1]{\@@startlink{#1}\@@href}%
\providecommand \@@href[1]{\endgroup#1\@@endlink}%
\providecommand \@sanitize@url [0]{\catcode `\\12\catcode `\$12\catcode
  `\&12\catcode `\#12\catcode `\^12\catcode `\_12\catcode `\%12\relax}%
\providecommand \@@startlink[1]{}%
\providecommand \@@endlink[0]{}%
\providecommand \url  [0]{\begingroup\@sanitize@url \@url }%
\providecommand \@url [1]{\endgroup\@href {#1}{\urlprefix }}%
\providecommand \urlprefix  [0]{URL }%
\providecommand \Eprint [0]{\href }%
\providecommand \doibase [0]{http://dx.doi.org/}%
\providecommand \selectlanguage [0]{\@gobble}%
\providecommand \bibinfo  [0]{\@secondoftwo}%
\providecommand \bibfield  [0]{\@secondoftwo}%
\providecommand \translation [1]{[#1]}%
\providecommand \BibitemOpen [0]{}%
\providecommand \bibitemStop [0]{}%
\providecommand \bibitemNoStop [0]{.\EOS\space}%
\providecommand \EOS [0]{\spacefactor3000\relax}%
\providecommand \BibitemShut  [1]{\csname bibitem#1\endcsname}%
\let\auto@bib@innerbib\@empty
\bibitem [{\citenamefont {Bustamante}\ \emph {et~al.}(2003)\citenamefont
  {Bustamante}, \citenamefont {Bryant},\ and\ \citenamefont
  {Smith}}]{bustamante2003}%
  \BibitemOpen
  \bibfield  {author} {\bibinfo {author} {\bibfnamefont {C.}~\bibnamefont
  {Bustamante}}, \bibinfo {author} {\bibfnamefont {Z.}~\bibnamefont {Bryant}},
  \ and\ \bibinfo {author} {\bibfnamefont {S.~B.}\ \bibnamefont {Smith}},\
  }\href@noop {} {\bibfield  {journal} {\bibinfo  {journal} {Nature}\ }\textbf
  {\bibinfo {volume} {421}},\ \bibinfo {pages} {423} (\bibinfo {year}
  {2003})}\BibitemShut {NoStop}%
\bibitem [{\citenamefont {Kellermayer}(2005)}]{kellermayer}%
  \BibitemOpen
  \bibfield  {author} {\bibinfo {author} {\bibfnamefont {M.~S.}\ \bibnamefont
  {Kellermayer}},\ }\href@noop {} {\bibfield  {journal} {\bibinfo  {journal}
  {Physiol. Meas.}\ }\textbf {\bibinfo {volume} {26}},\ \bibinfo {pages} {R119}
  (\bibinfo {year} {2005})}\BibitemShut {NoStop}%
\bibitem [{\citenamefont {Neuman}\ \emph {et~al.}(2007)\citenamefont {Neuman},
  \citenamefont {Lionnet},\ and\ \citenamefont {Allemand}}]{neuman}%
  \BibitemOpen
  \bibfield  {author} {\bibinfo {author} {\bibfnamefont {K.~C.}\ \bibnamefont
  {Neuman}}, \bibinfo {author} {\bibfnamefont {T.}~\bibnamefont {Lionnet}}, \
  and\ \bibinfo {author} {\bibfnamefont {J.-F.}\ \bibnamefont {Allemand}},\
  }\href@noop {} {\bibfield  {journal} {\bibinfo  {journal} {Annu. Rev. Mater.
  Res.}\ }\textbf {\bibinfo {volume} {37}},\ \bibinfo {pages} {33} (\bibinfo
  {year} {2007})}\BibitemShut {NoStop}%
\bibitem [{\citenamefont {Deniz}\ \emph {et~al.}(2005)\citenamefont {Deniz},
  \citenamefont {Mukhopadhyay},\ and\ \citenamefont {Lenke}}]{deniz}%
  \BibitemOpen
  \bibfield  {author} {\bibinfo {author} {\bibfnamefont {A.~A.}\ \bibnamefont
  {Deniz}}, \bibinfo {author} {\bibfnamefont {S.}~\bibnamefont {Mukhopadhyay}},
  \ and\ \bibinfo {author} {\bibfnamefont {E.~A.}\ \bibnamefont {Lenke}},\
  }\href@noop {} {\bibfield  {journal} {\bibinfo  {journal} {J. R. Soc.
  Interface}\ }\textbf {\bibinfo {volume} {5}},\ \bibinfo {pages} {15}
  (\bibinfo {year} {2005})}\BibitemShut {NoStop}%
\bibitem [{\citenamefont {Neuman}\ and\ \citenamefont
  {Nagy}(2008)}]{neuman2008}%
  \BibitemOpen
  \bibfield  {author} {\bibinfo {author} {\bibfnamefont {K.~C.}\ \bibnamefont
  {Neuman}}\ and\ \bibinfo {author} {\bibfnamefont {A.}~\bibnamefont {Nagy}},\
  }\href@noop {} {\bibfield  {journal} {\bibinfo  {journal} {Nature Methods}\
  }\textbf {\bibinfo {volume} {5}},\ \bibinfo {pages} {491} (\bibinfo {year}
  {2008})}\BibitemShut {NoStop}%
\bibitem [{\citenamefont {Fisher}\ \emph {et~al.}(1999)\citenamefont {Fisher},
  \citenamefont {Marszalek}, \citenamefont {Oberhauser}, \citenamefont
  {Carrion-Vazquez},\ and\ \citenamefont {Fernandez}}]{fisher}%
  \BibitemOpen
  \bibfield  {author} {\bibinfo {author} {\bibfnamefont {T.~E.}\ \bibnamefont
  {Fisher}}, \bibinfo {author} {\bibfnamefont {P.~E.}\ \bibnamefont
  {Marszalek}}, \bibinfo {author} {\bibfnamefont {A.~F.}\ \bibnamefont
  {Oberhauser}}, \bibinfo {author} {\bibfnamefont {M.}~\bibnamefont
  {Carrion-Vazquez}}, \ and\ \bibinfo {author} {\bibfnamefont {J.~M.}\
  \bibnamefont {Fernandez}},\ }\href@noop {} {\bibfield  {journal} {\bibinfo
  {journal} {J. Physiol.}\ }\textbf {\bibinfo {volume} {520}},\ \bibinfo
  {pages} {5} (\bibinfo {year} {1999})}\BibitemShut {NoStop}%
\bibitem [{\citenamefont {Kishino}\ and\ \citenamefont
  {Yanagida}(1988)}]{kishino}%
  \BibitemOpen
  \bibfield  {author} {\bibinfo {author} {\bibfnamefont {A.}~\bibnamefont
  {Kishino}}\ and\ \bibinfo {author} {\bibfnamefont {T.}~\bibnamefont
  {Yanagida}},\ }\href@noop {} {\bibfield  {journal} {\bibinfo  {journal}
  {Nature}\ }\textbf {\bibinfo {volume} {334}},\ \bibinfo {pages} {74}
  (\bibinfo {year} {1988})}\BibitemShut {NoStop}%
\bibitem [{\citenamefont {Neuman}\ and\ \citenamefont
  {Block}(2004)}]{neuman2004}%
  \BibitemOpen
  \bibfield  {author} {\bibinfo {author} {\bibfnamefont {K.}~\bibnamefont
  {Neuman}}\ and\ \bibinfo {author} {\bibfnamefont {S.}~\bibnamefont {Block}},\
  }\href@noop {} {\bibfield  {journal} {\bibinfo  {journal} {Rev. Sci.
  Instrum.}\ }\textbf {\bibinfo {volume} {75}},\ \bibinfo {pages} {2787}
  (\bibinfo {year} {2004})}\BibitemShut {NoStop}%
\bibitem [{\citenamefont {Horme\~{n}o}\ and\ \citenamefont
  {Arias-Gonzalez}(2006)}]{hormeno}%
  \BibitemOpen
  \bibfield  {author} {\bibinfo {author} {\bibfnamefont {S.}~\bibnamefont
  {Horme\~{n}o}}\ and\ \bibinfo {author} {\bibfnamefont {J.~R.}\ \bibnamefont
  {Arias-Gonzalez}},\ }\href@noop {} {\bibfield  {journal} {\bibinfo  {journal}
  {Biol. Cell}\ }\textbf {\bibinfo {volume} {98}},\ \bibinfo {pages} {679}
  (\bibinfo {year} {2006})}\BibitemShut {NoStop}%
\bibitem [{\citenamefont {Gosse}\ and\ \citenamefont
  {Croquette}(2002)}]{gosse}%
  \BibitemOpen
  \bibfield  {author} {\bibinfo {author} {\bibfnamefont {S.}~\bibnamefont
  {Gosse}}\ and\ \bibinfo {author} {\bibfnamefont {V.}~\bibnamefont
  {Croquette}},\ }\href@noop {} {\bibfield  {journal} {\bibinfo  {journal}
  {Biophys. J.}\ }\textbf {\bibinfo {volume} {82}},\ \bibinfo {pages} {3314}
  (\bibinfo {year} {2002})}\BibitemShut {NoStop}%
\bibitem [{\citenamefont {Kikuchi}\ \emph {et~al.}(1997)\citenamefont
  {Kikuchi}, \citenamefont {Yokoyama},\ and\ \citenamefont
  {Kajiyama}}]{kikuchi1997}%
  \BibitemOpen
  \bibfield  {author} {\bibinfo {author} {\bibfnamefont {H.}~\bibnamefont
  {Kikuchi}}, \bibinfo {author} {\bibfnamefont {N.}~\bibnamefont {Yokoyama}}, \
  and\ \bibinfo {author} {\bibfnamefont {T.}~\bibnamefont {Kajiyama}},\
  }\href@noop {} {\bibfield  {journal} {\bibinfo  {journal} {Chemistry
  Letters}\ }\textbf {\bibinfo {volume} {26}},\ \bibinfo {pages} {1107}
  (\bibinfo {year} {1997})}\BibitemShut {NoStop}%
\bibitem [{\citenamefont {Bustamante}\ \emph {et~al.}(1997)\citenamefont
  {Bustamante}, \citenamefont {Rivetti},\ and\ \citenamefont
  {Keller}}]{brk1997}%
  \BibitemOpen
  \bibfield  {author} {\bibinfo {author} {\bibfnamefont {C.}~\bibnamefont
  {Bustamante}}, \bibinfo {author} {\bibfnamefont {C.}~\bibnamefont {Rivetti}},
  \ and\ \bibinfo {author} {\bibfnamefont {D.}~\bibnamefont {Keller}},\
  }\href@noop {} {\bibfield  {journal} {\bibinfo  {journal} {Curr. Opin.
  Struct. Biol.}\ }\textbf {\bibinfo {volume} {7}},\ \bibinfo {pages} {709}
  (\bibinfo {year} {1997})}\BibitemShut {NoStop}%
\bibitem [{\citenamefont {Drake}\ \emph {et~al.}(1989)\citenamefont {Drake}
  \emph {et~al.}}]{drake1989}%
  \BibitemOpen
  \bibfield  {author} {\bibinfo {author} {\bibfnamefont {B.}~\bibnamefont
  {Drake}} \emph {et~al.},\ }\href@noop {} {\bibfield  {journal} {\bibinfo
  {journal} {Science}\ }\textbf {\bibinfo {volume} {243}},\ \bibinfo {pages}
  {1586} (\bibinfo {year} {1989})}\BibitemShut {NoStop}%
\bibitem [{\citenamefont {de~Gennes}(1979)}]{degennesSC}%
  \BibitemOpen
  \bibfield  {author} {\bibinfo {author} {\bibfnamefont {P.-G.}\ \bibnamefont
  {de~Gennes}},\ }\href@noop {} {\emph {\bibinfo {title} {Scaling Concepts in
  Polymer Physics}}}\ (\bibinfo  {publisher} {Cornell University Press},\
  \bibinfo {address} {Ithaca, New York},\ \bibinfo {year} {1979})\BibitemShut
  {NoStop}%
\bibitem [{\citenamefont {Eisenriegler}\ \emph {et~al.}(1982)\citenamefont
  {Eisenriegler}, \citenamefont {Kremer},\ and\ \citenamefont
  {Binder}}]{Eisenriegler1982}%
  \BibitemOpen
  \bibfield  {author} {\bibinfo {author} {\bibfnamefont {E.}~\bibnamefont
  {Eisenriegler}}, \bibinfo {author} {\bibfnamefont {K.}~\bibnamefont
  {Kremer}}, \ and\ \bibinfo {author} {\bibfnamefont {K.}~\bibnamefont
  {Binder}},\ }\href@noop {} {\bibfield  {journal} {\bibinfo  {journal} {J.
  Chem. Phys.}\ }\textbf {\bibinfo {volume} {77}},\ \bibinfo {pages} {6296}
  (\bibinfo {year} {1982})}\BibitemShut {NoStop}%
\bibitem [{\citenamefont {Duplantier}\ and\ \citenamefont
  {Saleur}(1986)}]{Duplantier1986}%
  \BibitemOpen
  \bibfield  {author} {\bibinfo {author} {\bibfnamefont {B.}~\bibnamefont
  {Duplantier}}\ and\ \bibinfo {author} {\bibfnamefont {H.}~\bibnamefont
  {Saleur}},\ }\href {\doibase 10.1103/PhysRevLett.57.3179} {\bibfield
  {journal} {\bibinfo  {journal} {Phys. Rev. Lett.}\ }\textbf {\bibinfo
  {volume} {57}},\ \bibinfo {pages} {3179} (\bibinfo {year}
  {1986})}\BibitemShut {NoStop}%
\bibitem [{\citenamefont {Rowghanian}\ and\ \citenamefont
  {Grosberg}(2011)}]{Rowghanian2011}%
  \BibitemOpen
  \bibfield  {author} {\bibinfo {author} {\bibfnamefont {P.}~\bibnamefont
  {Rowghanian}}\ and\ \bibinfo {author} {\bibfnamefont {A.~Y.}\ \bibnamefont
  {Grosberg}},\ }\href@noop {} {\bibfield  {journal} {\bibinfo  {journal} {J.
  Phys. Chem. B}\ }\textbf {\bibinfo {volume} {in print}} (\bibinfo {year}
  {2011})}\BibitemShut {NoStop}%
\bibitem [{\citenamefont {Bubis}\ \emph {et~al.}(2009)\citenamefont {Bubis},
  \citenamefont {Kantor},\ and\ \citenamefont {Kardar}}]{bubis}%
  \BibitemOpen
  \bibfield  {author} {\bibinfo {author} {\bibfnamefont {R.}~\bibnamefont
  {Bubis}}, \bibinfo {author} {\bibfnamefont {Y.}~\bibnamefont {Kantor}}, \
  and\ \bibinfo {author} {\bibfnamefont {M.}~\bibnamefont {Kardar}},\
  }\href@noop {} {\bibfield  {journal} {\bibinfo  {journal} {Europhys. Lett.}\
  }\textbf {\bibinfo {volume} {88}},\ \bibinfo {pages} {48001} (\bibinfo {year}
  {2009})}\BibitemShut {NoStop}%
\bibitem [{\citenamefont {Ben-Naim}\ and\ \citenamefont
  {Krapivsky}(2010)}]{BenNaim}%
  \BibitemOpen
  \bibfield  {author} {\bibinfo {author} {\bibfnamefont {E.}~\bibnamefont
  {Ben-Naim}}\ and\ \bibinfo {author} {\bibfnamefont {P.~L.}\ \bibnamefont
  {Krapivsky}},\ }\href@noop {} {\bibfield  {journal} {\bibinfo  {journal} {J.
  Phys. A: Math. Theor.}\ }\textbf {\bibinfo {volume} {43}},\ \bibinfo {pages}
  {495007} (\bibinfo {year} {2010})}\BibitemShut {NoStop}%
\bibitem [{\citenamefont {Redner}(1983)}]{redner_book}%
  \BibitemOpen
  \bibfield  {author} {\bibinfo {author} {\bibfnamefont {S.}~\bibnamefont
  {Redner}},\ }\href@noop {} {\emph {\bibinfo {title} {A Guide to First-Passage
  Processes}}}\ (\bibinfo  {publisher} {Cambridge University Press},\ \bibinfo
  {address} {New York},\ \bibinfo {year} {1983})\BibitemShut {NoStop}%
\bibitem [{\citenamefont {Binder}(1983)}]{binder}%
  \BibitemOpen
  \bibfield  {author} {\bibinfo {author} {\bibfnamefont {K.}~\bibnamefont
  {Binder}},\ }in\ \href@noop {} {\emph {\bibinfo {booktitle} {Phase
  Transitions and Critical Phenomena}}},\ Vol.~\bibinfo {volume} {8},\ \bibinfo
  {editor} {edited by\ \bibinfo {editor} {\bibfnamefont {C.}~\bibnamefont
  {Domb}}\ and\ \bibinfo {editor} {\bibfnamefont {J.~L.}\ \bibnamefont
  {Lebowitz}}}\ (\bibinfo  {publisher} {Academic Press},\ \bibinfo {address}
  {London},\ \bibinfo {year} {1983})\ p.~\bibinfo {pages} {1}\BibitemShut
  {NoStop}%
\bibitem [{\citenamefont {Caracciolo}\ \emph {et~al.}(1998)\citenamefont
  {Caracciolo}, \citenamefont {Causo},\ and\ \citenamefont
  {Pelissetto}}]{caracciolo}%
  \BibitemOpen
  \bibfield  {author} {\bibinfo {author} {\bibfnamefont {S.}~\bibnamefont
  {Caracciolo}}, \bibinfo {author} {\bibfnamefont {M.~S.}\ \bibnamefont
  {Causo}}, \ and\ \bibinfo {author} {\bibfnamefont {A.}~\bibnamefont
  {Pelissetto}},\ }\href@noop {} {\bibfield  {journal} {\bibinfo  {journal}
  {Phys. Rev. E}\ }\textbf {\bibinfo {volume} {57}},\ \bibinfo {pages} {R1215}
  (\bibinfo {year} {1998})}\BibitemShut {NoStop}%
\bibitem [{\citenamefont {Madras}\ and\ \citenamefont
  {Slade}(1993)}]{madras_book}%
  \BibitemOpen
  \bibfield  {author} {\bibinfo {author} {\bibfnamefont {N.}~\bibnamefont
  {Madras}}\ and\ \bibinfo {author} {\bibfnamefont {G.}~\bibnamefont {Slade}},\
  }\href@noop {} {\emph {\bibinfo {title} {The Self-Avoiding Walk}}}\ (\bibinfo
   {publisher} {Birkh{\"a}user},\ \bibinfo {address} {Boston},\ \bibinfo {year}
  {1993})\BibitemShut {NoStop}%
\bibitem [{\citenamefont {Cardy}\ and\ \citenamefont
  {Redner}(1984)}]{cardy_red}%
  \BibitemOpen
  \bibfield  {author} {\bibinfo {author} {\bibfnamefont {J.~L.}\ \bibnamefont
  {Cardy}}\ and\ \bibinfo {author} {\bibfnamefont {S.}~\bibnamefont {Redner}},\
  }\href@noop {} {\bibfield  {journal} {\bibinfo  {journal} {J. Phys. A}\
  }\textbf {\bibinfo {volume} {17}},\ \bibinfo {pages} {L933} (\bibinfo {year}
  {1984})}\BibitemShut {NoStop}%
\bibitem [{\citenamefont {Cardy}(1984)}]{cardy1}%
  \BibitemOpen
  \bibfield  {author} {\bibinfo {author} {\bibfnamefont {J.~L.}\ \bibnamefont
  {Cardy}},\ }\href@noop {} {\bibfield  {journal} {\bibinfo  {journal} {Nucl.
  Phys. B}\ }\textbf {\bibinfo {volume} {240}},\ \bibinfo {pages} {514}
  (\bibinfo {year} {1984})}\BibitemShut {NoStop}%
\bibitem [{\citenamefont {Cardy}(1983)}]{cardy2}%
  \BibitemOpen
  \bibfield  {author} {\bibinfo {author} {\bibfnamefont {J.~L.}\ \bibnamefont
  {Cardy}},\ }\href@noop {} {\bibfield  {journal} {\bibinfo  {journal} {J.
  Phys. A: Math. Gen.}\ }\textbf {\bibinfo {volume} {16}},\ \bibinfo {pages}
  {3617} (\bibinfo {year} {1983})}\BibitemShut {NoStop}%
\bibitem [{\citenamefont {Gutmann}\ and\ \citenamefont
  {Torrie}(1984)}]{guttmann}%
  \BibitemOpen
  \bibfield  {author} {\bibinfo {author} {\bibfnamefont {A.~J.}\ \bibnamefont
  {Gutmann}}\ and\ \bibinfo {author} {\bibfnamefont {G.~M.}\ \bibnamefont
  {Torrie}},\ }\href@noop {} {\bibfield  {journal} {\bibinfo  {journal} {J.
  Phys. A: Math. Gen.}\ }\textbf {\bibinfo {volume} {17}},\ \bibinfo {pages}
  {3539} (\bibinfo {year} {1984})}\BibitemShut {NoStop}%
\bibitem [{\citenamefont {Barber}\ \emph {et~al.}(1978)\citenamefont {Barber},
  \citenamefont {Guttmann}, \citenamefont {Middlemiss}, \citenamefont
  {Torrie},\ and\ \citenamefont {Whittington}}]{gsurface}%
  \BibitemOpen
  \bibfield  {author} {\bibinfo {author} {\bibfnamefont {M.~N.}\ \bibnamefont
  {Barber}}, \bibinfo {author} {\bibfnamefont {A.~J.}\ \bibnamefont
  {Guttmann}}, \bibinfo {author} {\bibfnamefont {K.~M.}\ \bibnamefont
  {Middlemiss}}, \bibinfo {author} {\bibfnamefont {G.~M.}\ \bibnamefont
  {Torrie}}, \ and\ \bibinfo {author} {\bibfnamefont {S.~G.}\ \bibnamefont
  {Whittington}},\ }\href@noop {} {\bibfield  {journal} {\bibinfo  {journal}
  {J. Phys. A}\ }\textbf {\bibinfo {volume} {11}},\ \bibinfo {pages} {1833}
  (\bibinfo {year} {1978})}\BibitemShut {NoStop}%
\bibitem [{\citenamefont {De'Bell}\ and\ \citenamefont
  {Lookman}(1993)}]{debell}%
  \BibitemOpen
  \bibfield  {author} {\bibinfo {author} {\bibfnamefont {K.}~\bibnamefont
  {De'Bell}}\ and\ \bibinfo {author} {\bibfnamefont {T.}~\bibnamefont
  {Lookman}},\ }\href@noop {} {\bibfield  {journal} {\bibinfo  {journal} {Rev.
  Mod. Phys.}\ }\textbf {\bibinfo {volume} {65}},\ \bibinfo {pages} {87}
  (\bibinfo {year} {1993})}\BibitemShut {NoStop}%
\bibitem [{\citenamefont {Slutsky}\ \emph {et~al.}(2005)\citenamefont
  {Slutsky}, \citenamefont {Zandi}, \citenamefont {Kantor},\ and\ \citenamefont
  {Kardar}}]{slutsky}%
  \BibitemOpen
  \bibfield  {author} {\bibinfo {author} {\bibfnamefont {M.}~\bibnamefont
  {Slutsky}}, \bibinfo {author} {\bibfnamefont {R.}~\bibnamefont {Zandi}},
  \bibinfo {author} {\bibfnamefont {Y.}~\bibnamefont {Kantor}}, \ and\ \bibinfo
  {author} {\bibfnamefont {M.}~\bibnamefont {Kardar}},\ }\href@noop {}
  {\bibfield  {journal} {\bibinfo  {journal} {Phys. Rev. Lett.}\ }\textbf
  {\bibinfo {volume} {94}},\ \bibinfo {pages} {198303} (\bibinfo {year}
  {2005})}\BibitemShut {NoStop}%
\bibitem [{\citenamefont {Kosmas}(1985)}]{kosmas}%
  \BibitemOpen
  \bibfield  {author} {\bibinfo {author} {\bibfnamefont {M.~K.}\ \bibnamefont
  {Kosmas}},\ }\href@noop {} {\bibfield  {journal} {\bibinfo  {journal} {J.
  Phys. A}\ }\textbf {\bibinfo {volume} {18}},\ \bibinfo {pages} {539}
  (\bibinfo {year} {1985})}\BibitemShut {NoStop}%
\bibitem [{\citenamefont {Douglas}\ and\ \citenamefont
  {Kosmas}(1989)}]{douglas_kosmas}%
  \BibitemOpen
  \bibfield  {author} {\bibinfo {author} {\bibfnamefont {J.~F.}\ \bibnamefont
  {Douglas}}\ and\ \bibinfo {author} {\bibfnamefont {M.~K.}\ \bibnamefont
  {Kosmas}},\ }\href@noop {} {\bibfield  {journal} {\bibinfo  {journal}
  {Macromolecules}\ }\textbf {\bibinfo {volume} {22}},\ \bibinfo {pages} {2412}
  (\bibinfo {year} {1989})}\BibitemShut {NoStop}%
\bibitem [{\citenamefont {Grassberger}(2005)}]{grass_gamma1}%
  \BibitemOpen
  \bibfield  {author} {\bibinfo {author} {\bibfnamefont {P.}~\bibnamefont
  {Grassberger}},\ }\href@noop {} {\bibfield  {journal} {\bibinfo  {journal}
  {J. Phys. A: Math. Gen.}\ }\textbf {\bibinfo {volume} {38}},\ \bibinfo
  {pages} {323} (\bibinfo {year} {2005})}\BibitemShut {NoStop}%
\bibitem [{\citenamefont {Suzuki}(1968)}]{dimerization1}%
  \BibitemOpen
  \bibfield  {author} {\bibinfo {author} {\bibfnamefont {K.}~\bibnamefont
  {Suzuki}},\ }\href@noop {} {\bibfield  {journal} {\bibinfo  {journal} {Bull.
  Chem. Soc. Japan}\ }\textbf {\bibinfo {volume} {41}},\ \bibinfo {pages} {538}
  (\bibinfo {year} {1968})}\BibitemShut {NoStop}%
\bibitem [{\citenamefont {Alexandrowicz}(1969)}]{dimerization2}%
  \BibitemOpen
  \bibfield  {author} {\bibinfo {author} {\bibfnamefont {Z.}~\bibnamefont
  {Alexandrowicz}},\ }\href@noop {} {\bibfield  {journal} {\bibinfo  {journal}
  {J. Chem. Phys.}\ }\textbf {\bibinfo {volume} {51}},\ \bibinfo {pages} {561}
  (\bibinfo {year} {1969})}\BibitemShut {NoStop}%
\bibitem [{\citenamefont {Maghrebi}\ \emph {et~al.}(2011)\citenamefont
  {Maghrebi}, \citenamefont {Rahi}, \citenamefont {Emig}, \citenamefont
  {Graham}, \citenamefont {Jaffe},\ and\ \citenamefont {Kardar}}]{Maghrebi10}%
  \BibitemOpen
  \bibfield  {author} {\bibinfo {author} {\bibfnamefont {M.~F.}\ \bibnamefont
  {Maghrebi}}, \bibinfo {author} {\bibfnamefont {S.~J.}\ \bibnamefont {Rahi}},
  \bibinfo {author} {\bibfnamefont {T.}~\bibnamefont {Emig}}, \bibinfo {author}
  {\bibfnamefont {N.}~\bibnamefont {Graham}}, \bibinfo {author} {\bibfnamefont
  {R.~L.}\ \bibnamefont {Jaffe}}, \ and\ \bibinfo {author} {\bibfnamefont
  {M.}~\bibnamefont {Kardar}},\ }\href {\doibase 10.1073/pnas.1018079108}
  {\bibfield  {journal} {\bibinfo  {journal} {PNAS}\ }\textbf {\bibinfo
  {volume} {108}},\ \bibinfo {pages} {6867} (\bibinfo {year}
  {2011})}\BibitemShut {NoStop}%
\bibitem [{\citenamefont {Ohno}(1989)}]{Ohno1989}%
  \BibitemOpen
  \bibfield  {author} {\bibinfo {author} {\bibfnamefont {K.}~\bibnamefont
  {Ohno}},\ }\href {\doibase 10.1103/PhysRevA.40.1524} {\bibfield  {journal}
  {\bibinfo  {journal} {Phys. Rev. A}\ }\textbf {\bibinfo {volume} {40}},\
  \bibinfo {pages} {1524} (\bibinfo {year} {1989})}\BibitemShut {NoStop}%
\end{thebibliography}
\end{document}